\documentclass{svproc}
\usepackage{url}
\usepackage{epsfig,amsmath,amssymb}

\begin{document}
\mainmatter              
\title{Optimal renormalization and the extraction of strange quark mass from semi-leptonic $\tau$-decay\thanks{Talk presented at the XXII DAE-BRNS High Energy Physics Symposium, Delhi University, Dec. 12-16, 2016.}}
\titlerunning{Extraction of the strange quark mass}  
%
\author{B. Ananthanarayan\inst{1} \and Diganta Das\inst{2}\thanks{Speaker.}}
\authorrunning{B. Ananthanarayan \emph{et al.}} 
%
%
\institute{Centre for High Energy Physics,
Indian Institute of Science, Bangalore 560 012, India,\\
\email{anant@chep.iisc.ernet.in}
\and
Physical Research Laboratory,  Navrangpura, Ahmedabad 380 009, India,\\
\email{diganta@prl.res.in}}

\maketitle              

\begin{abstract}
We employ optimal renormalization group analysis to semi-leptonic $\tau$-decay polarization functions and extract
the strange quark mass from their moments measured by the ALEPH and OPAL collaborations.
The optimal renormalization group makes use of the renormalization group equation of a given perturbation series which then leads
to closed form sum of all the renormalization group-accessible logarithms which have reduced scale
dependence. Using the latest theoretical inputs we find $m_s(2{\rm GeV}) = 106.70 \pm 9.36~{\rm MeV}$ and $m_s(2{\rm GeV}) = 74.47 \pm 7.77~{\rm MeV}$
for ALEPH and OPAL data respectively.
\keywords{renormalization group, strange quark mass, $\tau$-decay}
\end{abstract}
\section{Introduction}
The relevant quantities to extract strange quark mass $m_s$ from semi-leptonic $\tau$-decay are the polarization functions, the data on which
were made available some years ago by ALEPH \cite{Barate:1997hv} and OPAL \cite{Abbiendi:2004xa} collaborations.
One of the important theoretical objective in these extractions is to account for the renormalization group (RG) running of all the parameters that enter
the evolutions. There have been several schemes, for example \emph{fixed-order
perturbation theory} (FOPT) \cite{Jamin:2005ip}, \emph{contour-improved perturbation theory} (CIPT) 
\cite{Pivovarov:1991bj} and \emph{method of effective charges} (MEC) \cite{Groote:1997kh} to account for these runnings. 
We follow the optimal renormalization group analysis \cite{Maxwell:1999dv,Ahmady:2002fd} in Ref.~\cite{Ananthanarayan:2016kll} that uses the RG constraints
of a given perturbation series to obtain a closed form sum all the RG-accessible logarithms which substantially reduce the RG scale ($\mu$) dependence.
The RG-accessible logarithms are defined as all the leading and the next-to-leading logarithms that can be accessed through RG equation.
We call this scheme \emph{renormalization group summed perturbation theory} (RGSPT). 

After a brief summary of the formalism of 
semi-leptonic $\tau$-decay in Sec.~\ref{form}, we describe the derivations of the closed form summation in Sec.~\ref{rgsum}.
The extraction of $m_s$ is described in Sec.~\ref{ms} and we conclude in Sec. \ref{summary}.

\section{Formalism \label{form}}
The experimentally measurable quantities relevant for the extraction of $m_s$ from Cabibbo suppressed semi-leptonic $\tau$-decay 
are the moments of polarization functions. These can be written as a contour integral of the polarization
functions with suitable weight functions in the complex $q^2$ plane running counter clockwise along the circle $|q^2|=M_\tau^2$
\begin{eqnarray}\label{eq:Rkl2}
R^{kl}_{\tau} &=&  \frac{6i}{2\pi} \oint \rho_{k,l}(q^2) \frac{m^2_s}{q^2} \Pi^{mq} \frac{dq^2}{M_\tau^2} 
 - \frac{6i}{2\pi} \oint \Bigg[ \frac{m_s^2}{q^2}D^{mg} \int \rho_{k,l}(q^2) \frac{dq^2}{M_\tau^2}  \Big] \frac{dq^2}{M_\tau^2} \, .
\end{eqnarray}
Here $\rho_{k,l}$ are the weight functions, $\Pi^{mq}$ is the $\tau$-decay polarization function and $D^{mg}$ is the Adler function related to the polarization function $\Pi^{mg}$
and defined as $D^{mg} = -Q^2\frac{d\Pi^{mg}}{dQ^2}$, where $Q^2=-q^2$. The functions $\Pi^{mq}$ \cite{Chetyrkin:1993hi,Baikov:2004tk}  
and $D^{mg}$ \cite{Baikov:2002uw} are known to order $\alpha_s^3$ in the perturbation theory, and they have in addition to $Q^2$ dependence,
$\mu$ dependence through terms like $\alpha_s^n(\mu^2)\ln^{n-k}(\mu^2/Q^2)$,
at each order $n$ in the perturbation expansion. Both $\Pi^{mq}$ and $D^{mg}$ satisfy homogeneous RG equations.
In the next section we show how the RG equations of these functions can be used to obtain closed
form sums of all the RG accessible logarithms which reduce $\mu$ dependence. 


\section{Closed form sum RG-accessible logarithms \label{rgsum}}
We demonstrate the derivations of RGSPT series with $D^{mg}$. We write down the un-summed and RGSPT 
series as $D^{mg} = \sum_{n=0}^\infty \sum_{k=0}^n d^{mg}_{n,k} a_s^n L^k$ and 
$D^{mg}_{{\rm RGSPT}} = \sum_{n=0}^{\infty} a_s^n \Pi^{mg}_n[a_s L]$, respectively.
Here the intermediate quantities are defined as $D_k^{mg}[a_s L] = \sum_{n=k}^\infty d^{mg}_{n,n-k}(a_s L)^{n-k}$, $d^{mg}$ 
are the series coefficients that can be extracted from the expression of $D^{mg}$ given in the appendix of
Ref.~\cite{Ananthanarayan:2016kll}, $L = \ln(\mu^2/Q^2)$, and $a_s=\alpha_s(\mu)/\pi$. 
We substitute the un-summed expression of $D^{mg}$ in the homogeneous RG equation that it satisfies and collect the coefficients of 
$a_s^n L^{n-1-k}$ which leads to a recursion relation in terms of the series coefficients
$d^{mg}$. The recursion relation is then multiplied by $(a_sL)^{n-1-k}$ and is summed from $n=1+k$ to infinity, 
which following the definitions of the intermediate quantities $D_k^{mg}[a_s L]$ results in differential
equations for them. These differential equations are solved with suitable boundary conditions
and the resultant solutions are the closed form expressions of $D^{mg}_k[a_s L]$. 
The closed form expressions of $D^{mg}_{k}[a_s L]$ are given in \cite{Ananthanarayan:2016kll}.
\section{Extraction of $m_s$ from ALEPH and OPAL data\label{ms}}
The extraction of $m_s$ is possible by the measurements of the $SU(3)$ flavor breaking terms 
$\delta R^{kl}_\tau = N_c S_{EW} \Big(-R^{kl} -4\pi^2 \frac{m_s^2(M_\tau^2)}{M_\tau} \frac{\langle\bar{s}s\rangle}{M_\tau^3}f_{kl} \Big)$
where the second term within the bracket is the contribution of the condensate \cite{Braaten:1991qm}. The vales of $S_{EW}, f_{kl}$ and 
$\langle \bar{s}s\rangle$ are collected in \cite{Ananthanarayan:2016kll}. The flavor breaking term $\delta R^{kl}_\tau$ for different moments $(k,l)$ have been
measured by ALEPH \cite{Barate:1997hv} and OPAL \cite{Abbiendi:2004xa} collaborations. For our determination, we have used the moments $(0,0), (1,0), (2,0)$ of ALEPH and $(2,0), (3,0)$
and $(4,0)$ of OPAL. The determinations for individual moments for both ALEPH and OPAL are tabulated in \cite{Ananthanarayan:2016kll} and are compared with the 
determinations in FOPT, CIPT and MEC schemes. We find that RGSPT is consistent with other schemes.
By doing a weighted average of the individual determinations we obtain the following final values of $m_s$ at the $\tau$ mass scale
%
$ m_s(M_\tau^2) = 110.18 \pm 9.67\, {\rm MeV} \, , m_s(M_\tau^2) = 76.90 \pm 8.03\, {\rm MeV}\, $
%
for ALEPH and OPAL data, respectively. Here the errors are dominated by experimental uncertainties.
Evolving these determinations to 2 GeV using \cite{Chetyrkin:2000yt} we get
\begin{equation}\label{eq:final2}
 m_s(2\text{GeV}) = 106.70 \pm 9.36\, {\rm MeV},\quad m_s(2\text{GeV}) = 74.47 \pm 7.77\,  {\rm MeV},
\end{equation}
for ALEPH and OPAL data, respectively. Within errors our determinations are comparable with that in lattice QCD.

\section{Summary and Conclusion \label{summary}}
We have applied for the first time the optimal renormalization group analysis to $\tau$-decay polarization functions in the context of extraction of strange
quark mass. Our determinations are consistent with other schemes and show that the strange quark mass extracted from $\tau$-decay is 
insensitive to the choice of renormalization schemes. Therefore, in the future when the experimental data improve, precise determinations
of $m_s$ can be made.

\end{document}